\documentstyle[12pt]{article}
\advance\oddsidemargin	by -1in
\advance\evensidemargin by -1in
\advance\topmargin	by -0.0in
\makeindex
\newcommand\beq{\begin{equation}}
\newcommand\eeq{\end{equation}}
\newcommand\beqn{\begin{eqnarray}}
\newcommand\eeqn{\end{eqnarray}}
\newcommand{\doublespace} {
    \renewcommand{\baselinestretch} {1.6} \large\normalsize}

\begin{document}
\vspace*{0.5cm}
\hspace*{9cm}{\Large\bf MPIH-V12-1995}
\vspace*{2cm}

\centerline{{\huge\bf Nuclear Shadowing and}}
\medskip
\centerline{{\huge \bf the Proton Structure Function}}
\medskip
\centerline{{\huge \bf	at Small x}}
\vspace{1.5cm}
\begin{center}
{\large Boris~Kopeliovich\footnote{On leave from
 the Joint
 Institute for
Nuclear Research, Laboratory of Nuclear Problems,
\newline Dubna, 141980 Moscow Region, Russia.
E-mail: bzk@dxnhd1.mpi-hd.mpg.de}$\ $
 and Bogdan~Povh}

\vspace{0.5cm}

 {\sl Max-Planck Institut f\"ur Kernphysik, Postfach 103980,
\newline 69029 Heidelberg, Germany}\\
\end{center}
\bigskip

\doublespace
\begin{abstract}

 A new scaling variable is introduced in terms of which nuclear
 shadowing
in deep-inelastic scattering is universal, i.e.
 independent of $A$, $Q^2$ and
$x$.  This variable can be
 interpreted as a measure of the number of gluons
probed by the
 hadronic fluctuations of a virtual photon during their
lifetime.
 According to recent data from the H1 and ZEUS experiments, the
gluon density in a proton rises steeply as $x$ tends to zero.  So
 nuclei
which in the infinite momentum frame have a large surface
 density of gluons
model at larger $x$ values what we expect for
 the proton at much smaller $x$.
Using experimental information
 on nuclear shadowing we predict the unitarity
corrections to the
 proton structure function at small $x$ and extract the
bare
 Pomeron intercept.
\\
\bigskip\\

\end{abstract}
\newpage
Nuclear shadowing at small Bjorken $x\ll 1$ is experimentally
 well
established in deep-inelastic lepton scattering (DIS)
\cite{nmc1,nmc2,e665} and in Drell-Yan lepton pair production
\cite{e772} .  Theoretically shadowing in the nuclear structure
functions has been predicted on the basis of the old fashioned
 parton
 model
\cite{kancheli,NZold} and more recently in QCD calculations
\cite{levin,mueller}. In both approaches shadowing is the result
of parton recombination.

 Shadowing, or unitarity corrections, is
supposed
 to exist in the proton structure function as well.  In this case,
however, there is no direct way to observe experimentally
 shadowing in the
proton because there is nothing to compare to.
 The unitarity corrections at
small $x$ are expected
 to be significant because the virtual photoabsorption
cross
 section rises steeply \cite{h1,zeus}, steeper than hadronic cross
sections at high energy. This fast growth is usually interpreted
 as being due
to the Pomeron contribution to the structure
 function corrected for
shadowing. No reliable way to calculate
 unitarity corrections has been
proposed.

 Since at small $x$ the parton clouds of different nucleons
overlap \cite{kancheli}, one expects larger parton densities, and
consequently stronger shadowing
 effects, in nuclei than in protons.  The
question arises: can
 one use the data on nuclear shadowing at small $x$ in
order
 to improve the reliability of the unitarity corrections to the
 proton
structure function? The aim of this letter is to
 answer this question.

\medskip

 There is a widespread belief that the qluon and
 sea quark distribution
functions experience nearly the same
 nuclear shadowing, because the sea,
which dominates at small $x$,
 is supposed to be generated through gluons.
This would be true,
 if our electromagnetic probes of the gluon distribution
were
 really hard. We argue below that this is not obvious even at high
$Q^2$.

 A photon, through its quark-antiquark fluctuations, interacts
with gluons.  This process can also be interpreted as an
 electromagnetic
interaction with gluonic $q\bar q$ fluctuations.
 At high $Q^2$ perturbative
QCD can be applied, and it is
 convenient to decompose the photon wave
function in a quark-gluon
 basis. Assuming that the transverse separation
$\rho$ of the
 $q\bar q$ fluctuations in a highly virtual photon is small
(see,
 however, the discussion below) and that the interaction cross
 section
vanishes as $\sigma(\rho) \propto \rho^2$
\cite{zkl,bbgg}, one can present the total photoabsorption cross
section in the following form \cite{barone,fs}

\beq
\sigma^{\gamma^*N}_{tot}(x,Q^2)\approx \frac{2\pi^2}
{3}\alpha_s(\rho)\ \langle\rho^2\rangle\ g_N(x,Q^2)
\label{1}
\eeq
Here $g_N(x,Q^2)=xG_N(x,Q^2)$ is the gluon distribution function.
 The
averaging in (\ref{1}) is weighted with the square of the
 wave
 function of
the $q\bar q$ fluctuation of the photon \cite{nz91},
$\langle\rho^2\rangle=\int_0^1d\alpha\int d^2\rho|\Psi_{\gamma^*}
^2(\rho,\alpha)|_T^2\ \rho^2$, where $\alpha$ is the fraction
 of the photon
light-cone momentum carried by the quark;
$|\Psi_{\gamma^*}^2(\rho,\alpha)|_T^2\propto
 [1-2\alpha(1-\alpha)]\
\epsilon^2K_1^2(\epsilon\rho)$ and
 $\epsilon^2=\alpha(1-\alpha)Q^2+m_q^2$ is
an important parameter
 which determines the transverse separation of the
$q\bar q$
 fluctuation, $\rho^2\propto 1/\epsilon^2$.  It is of order
$1/Q^2$ except at the edges of the kinematical region, where
 $\alpha$
 or
$1-\alpha\sim m_q^2/Q^2$, and the $q\bar q$ fluctuation
 acquires a large
transverse size, $\rho^2\sim 1/m_q^2$.
\cite{nz91,bk,fs1}.

 Of course such a component of the photon cannot be used as a
 probe of the
gluon distribution.  For instance, if one needs to
 estimate how many nucleons
are in a nucleus, and one uses, say,
 a pion-nucleus inelastic interaction as
a probe, one obviously
 will get a wrong result, i.e., that the number of
nucleons is
 $\sim A^{2/3}$.  The source of the trouble is the softness of
the
 probe, $\sigma(\pi N)$ is too large.  The same problem may arise
 in
DIS.  It is not clear what part of the observed nuclear
 shadowing comes from
the reduction of the gluon density caused by
 recombination, and what is due
to the softness of the
 $q\bar q$ probe.  The latter
 corresponds to the
multiple scattering shadowing mechanism.

 Although multiple scattering
corrections violate the
 factorization property of eq. (\ref{1}), it is
natural to assume
 that both shadowing mechanisms {\it solely} depend
 upon
the number of gluons involved in the
 interaction, rather than upon $x$, $Q^2$
or the nuclear
 mass number A. In this paper we will test this idea on
available data and use the results to predict shadowing for
 protons. We do
not try to disentangle the two sources of
 nuclear shadowing, mentioned above,
gluon recombination and
 multiple scattering.

 We begin by testing the
universality of the relation
 between the number of gluon participating in the
interaction and
 the ratio of the nucleus and nucleon structure
functions. The
 general expression for the nuclear photoabsorption
 cross
section \cite{kl78,zkl,nz91} reads

\beq
\sigma^{\gamma^*A}_{tot}(x,Q^2)=
2\int d^2b\left\langle 1-
\left[1- \frac{\sigma(\rho,x)T(b)}{2A}\right]^A
\right\rangle
\label{4}
\eeq
The averaging here is the same as defined in eq. (\ref{1}).
 $T(b)\approx
\int_{-\infty}^{\infty}dz\rho_A(b,z)$ is the nuclear thickness
function, where $\rho_A(b,z)$
 is the nuclear density, which dependents upon
the impact
 parameter $b$ and the
 longitudinal coordinate $z$.  The cross
section for
 the interaction of a $q\bar q$ fluctuation with a nucleon in
eq.
 (\ref{4}) can according to standard Regge phenomenology be
 represented
in the form,
 $\sigma(\rho,x)=\sigma(\rho)x^{-\Delta_P(Q^2)}$.

 Expanding
eq. (\ref{4}) one can
 represent nuclear shadowing effects in the form,

\beq
R_{A/N}(x,Q^2)=\frac{ \sigma^{\gamma^*A}_{tot}(x,Q^2)}
{A\sigma^{\gamma^*N}_{tot}(x,Q^2)}=
 1-{1\over
4}\frac{\langle\sigma^2(\rho)\rangle}
 {\langle\sigma(\rho)\rangle}
\langle T(b)\rangle\left({1\over x}\right)^{\Delta_P(Q^2)}
+...\ ,
\label{5}
\eeq
where $\langle T(b)\rangle=(A-1)/A^2\int d^2bT^2(b)$.

 Assuming
$\sigma(\rho)\approx C\rho^2$ we obtain

\beq
\frac{\langle\sigma^2(\rho)\rangle}
{\langle\sigma(\rho)\rangle}=
\frac{2.4C}{2m_q^2ln(Q^2/2m_q^2)}
\label{6}
\eeq

 Eq.  (\ref{5}) looks very similar to the well known Glauber
 approximation
\cite{glauber}.  It is, however, very
 different.  In the Glauber
approximation the first
 rescattering correction is proportional to
$\langle\sigma(\rho)\rangle\propto 1/Q^2$.  In our case it is
 proportional to
eq. (\ref{6}) and is of the order of
 $1/m_q^2$.  This comparison demonstrates
that, in terms of
 multiple scattering theory, nuclear shadowing in DIS is
dominated
 by inelastic shadowing \cite{gribov}, while the Glauber eikonal
contribution \cite{glauber} vanishes at high $Q^2$ \cite{nz91}.

 An
important ingredient of eq. (\ref{4}) is the assumption that
 the lifetime of
a photon fluctuation in the nuclear rest frame
 is long enough that it can be
regarded as propagating through the
 whole nucleus with
 a frozen intrinsic
separation $\rho$.  This is the same as saying
 that $x$ is sufficiently
small, $x\ll 1/m_NR^A$, tha all
 the parton clouds of nucleons with the same
impact
 parameter overlap in the infinite momentum frame
\cite{kancheli}.  However most available data are in the
transition region, where the lifetime, usually called
 the coherence
 time,
is comparable to the nuclear radius.  This can be taken
 into account by
introducing a phase shift between $q\bar q$ wave
 packets produced at
different longitudinal coordinates, in the
 same way as was done in refs.
\cite{neutrino,nz91}.  This is
 equivalent to the replacement of the mean
nuclear thickness
 function in eq.  (\ref{5}) by an effective one

\beq
\langle\widetilde{T}(b)\rangle={A-1\over A^2}\int d^2b
\left[\int_{-\infty}^{\infty}dz\ \rho_A(b,z)\ e^{iqz}\right]^2
\approx\langle T(b)\rangle\ F_A^2(q)\ .
\label{7}
\eeq

 We use a Gaussian form for the nuclear density in order to make a
factorization of expression (\ref{7}) possible.  Generally we use
 the
realistic nuclear density of ref.  \cite{density}.
 $F_A(q)=\exp(-q^2R_A^2/6)$
is the nuclear longitudinal
 formfactor, where $R_A^2$ is the mean square
nuclear radius.  The
 decrease of the effective nuclear thickness function at
large $q$
 can be interpreted as a result of the short path length of the
hadronic fluctuation in the nucleus if we sit in the latter's
 rest frame, or
as an incomplete overlap of the gluon clouds of
 the nucleons which have the
same impact parameter in the infinite
 momentum frame of the nucleus.

 In
order to calculate the longitudinal momentum transfer in DIS,
$q=(Q^2+M^2)/2\nu$, one needs to know the effective mass of the
 produced
$q\bar q$ wave packet.  However a $q\bar q$ state with
 definite separation
$\rho$ does not have a definite mass.  We
 evaluate
 $q\approx 2xm_N$
assuming $M^2\sim Q^2$.  Thus $x$ is a parameter
 which controls the value of
$\langle\widetilde{T}(b)\rangle$.

 According to eqs.  (\ref{5}) and
(\ref{7}) the number of gluons
 which interact with the $q\bar q$
fluctuation
 during its lifetime is

\beq
n(x,Q^2,A)={1\over 4}
\frac{\langle\sigma^2(\rho)\rangle}
{\langle\sigma(\rho)\rangle}
\langle T(b)\rangle F^2_A(q)
\left({1\over
x}\right)^{\Delta_P(Q^2)}\ .
\label{8}
\eeq

 We expect nuclear shadowing to scale as a function of this
 new variable
$n(x,Q^2,A)$. Now we are in position to test
 this prediction.

 First of
all we fixed $\Delta_P(Q^2)$ by a fit to recent data
 from the ZEUS
\cite{zeus} and H1 \cite{h1} experiments on the
 proton structure functions
using a parameterization
 $x^{-\Delta_P}$ at fixed $Q^2$.  The results of the
fit are
 plotted as circles (H1) and squares (ZEUS) versus $Q^2$ in Fig.
 1.
We also included a hadronic point which we fixed
 conventionally at
$\Delta_P(0)=0.1$ which corresponds to results
 from various fits of the
energy dependence of hadronic cross
 sections.

 The $Q^2$-dependence of
the $\Delta_P(Q^2)$ as shown in Fig.  1
 was fitted by the ansatz
$\alpha_P(Q^2)=a+b\exp(-cQ^2)$.  The
 solid curve corresponds to $a=0.36\pm
0.016$, $b=-0.26\pm
 0.016$ and $c=0.052\pm 0.009$.

 The expected
universal dependence of the nuclear shadowing on
 $n(x,Q^2,A)$ is not affected
by the overall normalization of this
 variable.  Provided that $Q^2\gg m_q^2$,
one can test this
 scaling unambiguously despite the uncertainty in the value
of the
 factor $C$ and the quark mass $m_q$ displayed in eq.  (\ref{6}).

The values of the variable $n(x,Q^2)$ have been calculated
 from eq.
(\ref{8}) using the data from the NMC
\cite{nmc1,nmc2} and E665 \cite{e665} experiments, as well as the
results of our fit to $\Delta_P(Q^2)$. The data on the ratio of
 the nuclear
and nucleonic photoabsorption cross sections
 $R_{A/N}(x,Q^2)$ is plotted
against the new variable
 $n(x,Q^2,A)$ in Fig.  2.  The data demonstrate an
excellent
 scaling in $n(x,Q^2,A)$.  The concrete choice of the parameters
$C$ and $m_q$ (which are strongly correlated) in eq.  (\ref{6})
 was
adjusted to fit eq. (\ref{5}),
 the solid curve, to the data. We fixed
$2m_q^2=0.1\ GeV^2$ and
 $C/2m_q^2=22$.

 We should comment on this
procedure:\\
 {\bf(i)} Our considerations are valid for small $x$, so we
limit
 the $x$-region to $x<0.07$. At this $x$ the nuclear structure
function shows a small enhancement relative to the proton one,
 what results
in $R_{A/N}(x,Q^2)>1$ for $n\ss 0$. For this reason
 we renormalized eq. (3)
by 3\%.

 {\bf(ii)} Data points \cite{nmc2,e665} for $Q^2<0.5\ GeV^2$,
were
 excluded from the analysis because they are in the realm of the
 vector
dominance model, rather than DIS.  They should correspond
 to the same nuclear
shadowing experienced by a $\rho$-meson as in
 the real photon limit. This is
the reason for the saturation of
 nuclear shadowing at small $x$, claimed in
\cite{e665,nmc2}.  On
 the contrary, nuclear shadowing in DIS at small $x$,
but high
 $Q^2$ is not supposed to saturate.  $R_{A/N}(x,Q^2)$ is expected
to decrease logarithmically down to small $x$ below the real
 photon limit due
to gluon fusion \cite{mueller,nz91,barone}.

 {\bf(iii)} The data in Fig.  2
show that $R_{A/N}(x,Q^2)$ depends
 to a good accuracy linearly upon
$n(x,Q^2,A)$ for $n<0.3$.  On
 the one hand, this implies that higher order
terms in the
 expansion in eq.  (\ref{5}) are small. On the other hand it
confirms the validity of the small-$\rho^2$ approximation for
 $\sigma(\rho)$
used in eq.  (\ref{1}) and eq.  (\ref{6}).  Data
 for heavier nuclei and/or
smaller $x$ will probably demand higher
 order terms in eq.  (\ref{5}).

\medskip

 The shadowing described by eq. (\ref{5}) depends entirely on the
 gluon
thickness function of the target.  It should, therefore, be
 possible to apply
the same formalism to the shadowing in DIS off
 protons.
 The nuclear
shadowing fixes the parameter $C/m_q^2$ of eq.
 (\ref{6}), and thus it reduces
ambiguities faced by our
 evaluation of the unitarity corrections to the
proton structure
 function.  For numerical calculations we used the mean
thickness
 function of proton $\langle T\rangle\ss 1/4\pi B$.  The value of
the slope parameter $B\approx 6\ GeV^{-2}$ was estimated using
 data on $pp$
elastic scattering, assuming the Pomeron
 factorization.

 The predicted
ratio of the measured proton structure function and
 the bare Pomeron
contribution, $F_2^p(x,Q^2)/[F_2^p(x,Q^2)]_P$,
 is presented in Fig.3 as
function of $x$ versus $Q^2$. For each
 value of $Q^2$ we have put constraints
on $n\leq 0.3$ where
 nuclear data exist and we are sure that the linear
approximation
 (\ref{5}) is valid.

 After subtraction of the unitarity
corrections to the proton
 structure function we can determine the bare
Pomeron intercept.
 We refitted the data \cite{h1,zeus} and deduce the new
values of
 $\Delta_P^0(Q^2)$ .  The new interpolating dashed curve
corresponds to the parameters defined above: $a=0.45\pm
 0.017$, $b=-0.35\pm
0.017$ and $c=.0525\pm 0.008$.  The
 function
 $\Delta_P^0(Q^2)$ which we
obtained is quite close at large $Q^2$
 to the intercept of the BFLK Pomeron
\cite{bflk}.

 Another assumption, which we quietly made above, should be
commented upon. As was
 mentioned above, there are two contributions to the
shadowing in
 DIS, one comes from the suppression of the gluon density as a
consequence of gluon fusion $gg\ss g$, which corresponds to the
 triple
Pomeron graph in the framework of standard Regge
 phenomenology. Another
contribution to the shadowing comes from
 the Glauber-like rescattering of the
$q\bar q$ fluctuation off
 gluons.  This process can also be represented as a
parton fusion,
 but as a fusion of gluons into a $q\bar q$ pair, $gg\ss q\bar
q$.
 In Regge-model language this process corresponds to the
Pomeron-Pomeron-Reggeon graph. Both mechanisms show the same
 $x$-dependence
$\propto 1/x^{2\Delta_P}$, except for logarithmic
 corrections.  The
logarithmic corrections for the triple-Pomeron
 contribution differ for nuclei
and the nucleon. Our estimate
 shows,
 however, that, in the $x$ and $A$
domain investigated here, this
 difference is small and it decreases with
$x\ss 0$.

 To summarize: we have found a new variable $n(x,Q^2,A)$ which
all available data on nuclear shadowing in DIS scale with at
 small-$x$.  This
variable gives a measure of the number of
 gluons which a hadronic fluctuation
of the virtual photon
 interacts with during its lifetime.

 The data on
nuclear shadowing allowed us to fix the important
 parameter $C/m_q^2$ which
determines the screening of the virtual
 photon, and this reduced the
uncertainties in the estimations of
 the unitarity corrections to the proton
structure function.  A
 value of the bare Pomeron intercept was evaluated from
the data
 on the proton structure function corrected for shadowing. This
intercept is stiil an effective one, valid in the $x$ region
 reached by
HERA.

 {\bf Acknowledgement:} We would like to thank J.~H\"ufner and
E.~Predazzi for useful discussions, D.~Jansen and M.~Lavelle for
 careful
reading of the manuscript and imprivement suggestions.
 B.K.  thanks MPI f\"ur
Kernphysik, Heidelberg, for financial
 support.

\bigskip

 {\bf Figure capture}

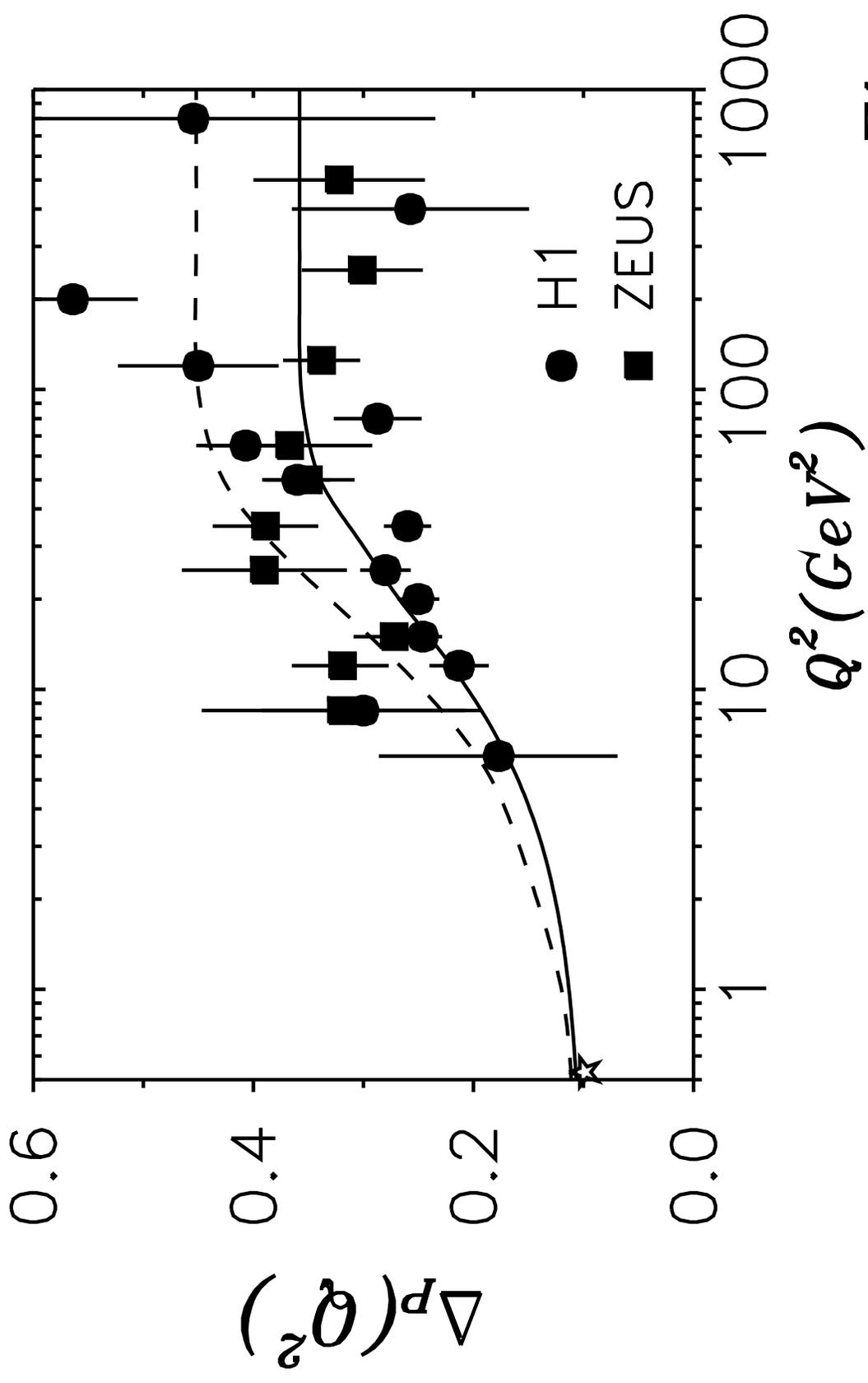
 {\bf Fig.  1} The Pomeron intercept in DIS off a
proton,
 $\alpha_P(Q^2)=1+\Delta_P(Q^2)$.  The data points are results of
 a
fit to the data on $x$-dependence of the proton structure
 function measured
by the H1 \cite{h1} and ZEUS \cite{zeus}
 experiments at HERA.  The solid
curve is the interpolation of the
 results of the fit to the effective
intercept. The dashed curve
 corresponds to the bare Pomeron contribution
after unitarity
 corrections are extracted from the data.

\medskip

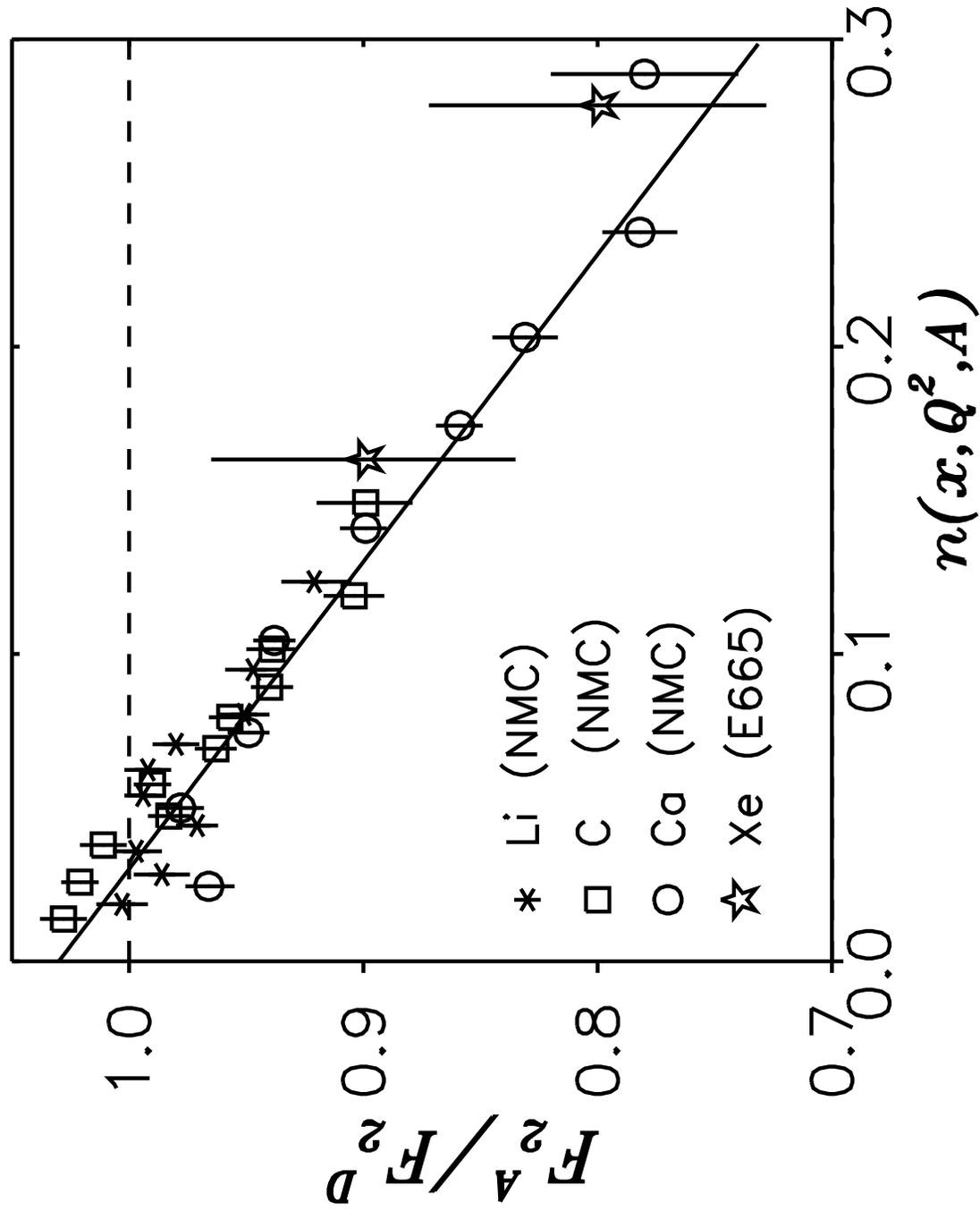
 {\bf Fig.  2} Data on nuclear shadowing at small $x$ from the NMC
\cite{nmc1,nmc2} and E665 \cite{e665} experiments versus the
scaling variable $n(x,Q^2,A)$ defined in eq.  (\ref{8}).  The
 straight line
corresponds to $F-2^A/F_2^D=1-n(x,Q^2,A)$.

\medskip

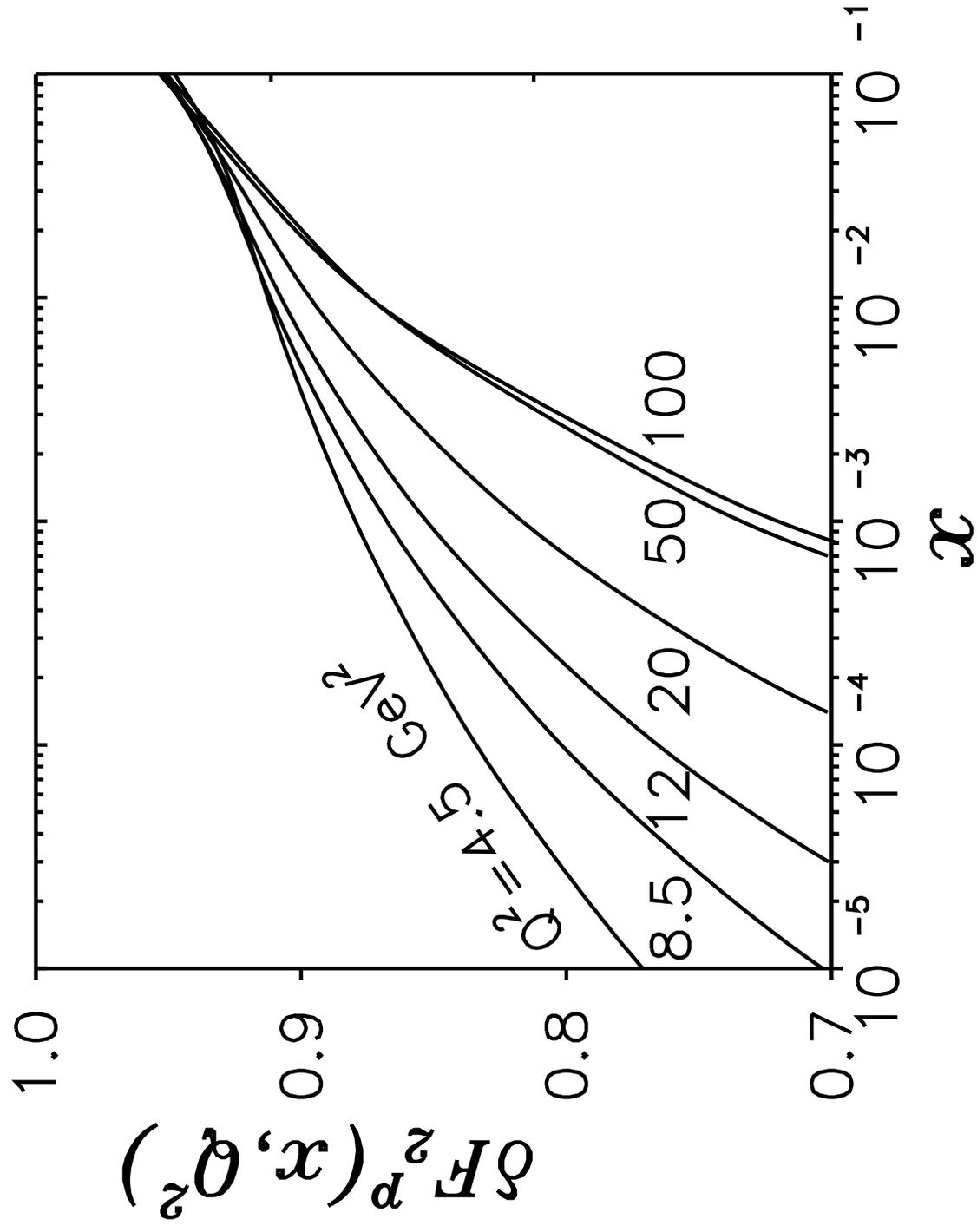
 {\bf Fig. 3} The unitarity/shadowing correction to the bare
 Pomeron
contribution to the proton structure function. Each curve
 corresponds to the
straight line in Fig. 2, plotted as a function
 of $x$ versus $Q^2$.


\begin{thebibliography}{MMM}
\baselineskip 10pt
\bibitem{nmc1} CERN NMC, P.~Amaudruz et al., Z.
Phys.  {\bf C51}, 387 (1991)
\bibitem{nmc2} CERN NMC, M.~Arneodo et al., LANL hep-ph/9504002,
Nucl. Phys. {\bf B 441}, 3 (1995)
\bibitem{e665} FNAL E665, M.R.~Adams et al., Phys. Lett.
{\bf B287}, 375 (1992)
\bibitem{e772} FNAL E772, E.~Alde et al., Phys. Rev. Lett. {\bf
64}, 2479 (1990)
\bibitem{kancheli} O.V.~Kancheli, Sov. Phys. JETP Lett. {\bf 18},
274 (1973)
\bibitem{NZold} N.N.~Nikolaev and V.I.~Zakharov, Phys. Lett. {\bf
B55}, 397 (1975)
\bibitem{levin} L.V.~Gribov, E.M.~Levin and M.G.~Ryskin, Phys.
Rept. {\bf 100}, 1 (1983); E.M.~Levin and M.G.~Ryskin, Sov. J.
Nucl. Phys. {\bf 41}, 300 (1985)
\bibitem{mueller} A.H.~Mueller and J.~Qiu, Nucl. Phys. {\bf
B268}, 427 (1986)
\bibitem{bflk} E.A.~Kuraev, L.N.~Lipatov and V.S.~Fadin,
Sov. Phys. JETP {\bf 44}, 443 (1976); {\bf 45}, 199 (1977);
 Ya.Ya.~Balitskii
and L.I.~Lipatov, Sov. J. Nucl. Phys. {\bf 28},
 822 (1978); L.N.~Lipatov,
Sov. Phys. JETP {\bf 63}, 904 (1986)
\bibitem{zeus} DESY ZEUS, M.~Derrick et al., DESY 94-143, August
1994
\bibitem{h1} DESY H1, T.~Ahmed et al., DESY 95-006, January 1995
\bibitem{zkl} Al.B.~Zamolodchikov, B.Z.~Kopeliovich
and L.I.~Lapidus, Sov. Phys. JETP Lett. {\bf 33}, 612 (1981)
\bibitem{bbgg}
G. Bertsch, S.J. Brodsky, A.S. Goldhaber and J.F. Gunion,
Phys. Rev. Lett. {\bf 47}, 297 (1981)
\bibitem{barone} V.~Barone et al., Z. Phys. {\bf C58}, 541 (1993)
\bibitem{fs} B.~Bl\"attel et al., Phys. Rev. Lett. {\bf
71}, 896 (1993)
\bibitem{bk} J.D.~Bjorken and J.~Kogut, Phys. Rev. {\bf D8}, 1341
(1973)
\bibitem{fs1} L.L.~Frankfurt and M.I.~Strikman, Phys. Rept.
{\bf 160}, 235 (1988)
\bibitem{nz91} N.N.~Nikolaev and B.G.~Zakharov, Z.  Phys.  {\bf
C49}, 607 (1991)
\bibitem{kl78} B.Z.~Kopeliovich and L.I.~Lapidus, Sov. Phys. JETP
Lett. {\bf 28}, 664 (1978)
\bibitem{glauber} R.J.~Glauber, In: Lectures in Theor. Phys., v.
1, ed. W.E.~Brittin and L.G.~Duham. NY: Intersciences, 1959
\bibitem{gribov} V.N.~Gribov, Sov. Phys. JETP {\bf 29}, 483
(1969)
\bibitem{neutrino} B.Z.~Kopeliovich, Phys. Lett. {\bf 227}, 461
(1989)
\bibitem{density} H.~De~Vries, C.W.~De Jager and C.~De~Vries,
Atomic Data and Nucl. Data Tables, {\bf 36}, 469 (1987)


\end{thebibliography}
\end{document}